\documentstyle[aps,prb,graphicx,amsbsy,amssymb]{revtex}
\begin{document}

\title{ON THE INFLUENCE OF A NON-LOCAL ELECTRODYNAMICS IN THE IRREVERSIBLE MAGNETIZATION OF NON-MAGNETIC BOROCARBIDES}
\author{A.V.~Silhanek, L.~Civale}
\address{Comisi\'{o}n Nacional de Energ\'{\i}a At\'{o}mica-Centro At\'{o}mico Bariloche\\ and Instituto Balseiro, 8400 Bariloche, Argentina.}
\author{J.R.~Thompson}
\address{Oak Ridge National Laboratory, Oak Ridge, Tennessee 37831-6061.\\Department of Physics, University of Tennessee, Knoxville, Tennessee 37996-1200.}
\author{P.C.~Canfield, S. L. Bud'ko}
\address{Ames Laboratory and Department of Physics and Astronomy, Iowa State University, Ames, Iowa 50011.}
\author{D.McK.~Paul}
\address{Department of Physics, University of Warwick, Coventry, CV4 7AL, United Kingdom.}
\author{C.V.~Tomy}
\address{Department of Physics, I.I.T. Powai, Mumbai, 400076, India.}



\maketitle

\date{\today}

\begin{abstract}
We present an overview of the temperature, field and angular dependence of the irreversible magnetization of non-magnetic borocarbides (Y;Lu)Ni$_2$B$_2$C. We show that nonlocal electrodynamics influences pinning via the unusual behavior of the shear modulus in non-hexagonal lattices. On top of that, we observe that the pinning force density $F_p$ exhibits a rich anisotropic behavior that sharply contrasts with its small mass anisotropy. When ${\bf H} \bot c$, $F_p$ is much larger and has a quite different $H$ dependence, indicating that other pinning mechanisms are present.

\end{abstract}

\section{INTRODUCTION}
Since early 1994, when the borocarbide family (\mbox{RNi$_2$B$_2$C}, where R = rare earth) was discovered\cite{mazumdar,nagarajan,cava94a,cava94b} a series of interesting results have been reported on these materials. The most remarkable feature is the coexistence of superconductivity and antiferromagnetism for R = Tm, Er, Ho and Dy. The  high superconducting transition temperatures $T_c$ and the broad variation of the ratio $T_N/T_c$ (where $T_N$ is the Neel temperature, ranging from $1.5$ K to $10$ K) make this family particularly appropriate to explore that coexistence\cite{yaron96,Eisaki94,Eskildsen98,Gammel99,Norgaard00}.

Interesting properties are also found in the non-magnetic borocarbides, R=Y,Lu. These materials exhibit a non-exponential temperature dependence\cite{carter,hong} and a non-linear field dependence\cite{nohara} of the electronic specific heat in the superconducting phase; an anomalous upper critical field $H_{c2}$ with an upward curvature near $T_c$\cite{shulga,ovchinnikov00} and a four-fold oscillation when the field is rotated within the basal plane\cite{metlushko97a}; and a square flux line lattice (FLL) at high fields\cite{yethiraj97a,mckpaul98}.

Although these properties were initially taken as evidence of a non conventional pairing mechanism\cite{wang-maki}, both the unconventional $H_{c2}$ and the square FLL can also be satisfactorily explained by assuming a non-local electrodynamics. Traditionally, nonlocality was expected to be relevant only in materials with a Ginzburg-Landau parameter $\kappa = \lambda / \xi \sim 1$ and sufficiently clean to have an electronic mean free path $\ell \gg \xi$. However, the large vortex cores make theoretical analysis very difficult in that case. The borocarbides, in contrast, have intermediate $\kappa$ values ($\sim 10 - 20$) that make core effects much smaller. This allows the non-local effects to be introduced perturbatively either in the London model, as done by Kogan, Gurevich et al.\cite{kogan96}, or in the Ginzburg-Landau theory, as shown by de Wilde et al.\cite{dewilde97} and Park and Huse\cite{park-huse}. In both cases, the result is a coupling of the supercurrents to the underlying crystal symmetry. 

The first success of the Kogan-Gurevich nonlocal model was the explanation\cite{kogan96} of the deviations of the reversible magnetization $M$ of Bi:2212 in the mixed state from the logarithmic dependence on magnetic field, $M \propto ln(H_{c2}/H)$, predicted by the London model\cite{kogan88a}. Later on, Song et al.\cite{song99a} found similar deviations in YNi$_2$B$_2$C when ${\bf H} \parallel c$-axis, that could also be quantitatively accounted for by the model. Recently we have extended that study to all directions of ${\bf H}$ and showed that this generalization of the London theory provides a satisfactory complete description of the anisotropic $M(H)$ with a self-consistent set of parameters\cite{basalplane}.

In the local London model anisotropy is introduced via a second rank mass tensor $m_{ij}$. In tetragonal materials such as borocarbides $m_a = m_b$, thus the properties should be isotropic in the basal plane. However, non-local corrections introduce\cite{kogan99a} a fourfold anisotropy as a function of the magnetic field orientation within the $ab$ plane, reflecting the square symmetry. This $\pi /2$ periodicity was indeed observed in $M$ in the mixed state of both YNi$_2$B$_2$C\cite{fourfold,physicaC} and LuNi$_2$B$_2$C\cite{kogan99a}.

Kogan's model also predicted\cite{kogan97a} that two structural transitions in the FLL should occur in borocarbides for ${\bf H} \parallel c$, a first order reorientation transition between two rhombic lattices at a field $H_1$ and a second order transition from rhombic to square at $H_2 > H_1$. Small angle neutron scattering (SANS) studies\cite{mckpaul98} confirmed those predictions in YNi$_2$B$_2$C. A jump in the apical angle $\beta$ of the rhombic lattice, discontinuous within the resolution, occurs at $H_1 \sim 1$ to $1.25kOe$, and the lattice becomes square ($\beta = 90^{\circ}$) at $H_2 \sim 1.25$ to $1.5kOe$. According to a recent analysis by Knigavko et al.\cite{knigavko}, $H_1$ really consists of two second order transitions taking place in a very narrow field range.

Although the role of nonlocality on the $\it {equilibrium}$ properties of the FLL is by now convincingly established, less is known about its effects on the $\it {nonequilibrium}$ vortex response. Since nonmagnetic borocarbides exhibit a very low critical current $J_c$ for ${\bf H} \parallel c$, pinning correlation volumes were expected to be large, as indeed observed\cite{eskildsen97b}. Thus, the elastic properties of the FLL must play a key role in the pinning. As the shear modulus $C_{66}$ depends on $\beta$\cite{rosenstein99}, and $\beta$ undergoes a discontinuous jump at $H_1$, it is bound to happen that $C_{66}$ and therefore the pinning properties change abruptly at this field\cite{eskildsen97b}. In other words, vortex pinning, which involves distortions from equilibrium vortex configurations, should be affected by the symmetry changes in the vortex lattice.

Recently we showed\cite{Fp} that, in YNi$_2$B$_2$C and for ${\bf H}\parallel c$, the reorientation transition at $H_1$ induces a kink in the pinning force density $F_p(H)$. We also observed anisotropies in $F_p$ both between the c-axis and the basal plane (out-of-plane anisotropy) and within the plane (in-plane anisotropy). We found that $F_p$ for ${\bf H} \bot c$ is one order of magnitude larger than for ${\bf H}\parallel c$ and has a quite different field dependence. We argued that this surprising behavior is unlikely to arise either from pinning by magnetic impurities or from non-local effects, although we recognized that the evidence supporting those claims was not conclusive.

Here we report further studies of $F_p$ in non-magnetic borocarbides. We show that the kink in $F_p(H)$ is also visible in LuNi$_2$B$_2$C and coincides with the field $H_1$ for this compound, thus confirming that it is a signature of this nonlocality-induced transition. We find that $H_1(T)$ slightly decreases as $T$ increases, in contrast to $H_2(T)$. We also study the effect of Co-doping in Lu(Ni$_{1-x}$Co$_x$)$_2$B$_2$C. We observe that $H_1$ decreases as the nonlocal effects are progressively suppressed by increasing $x$, in agreement with the $T$ dependence. All the LuNi$_{2-x}$Co$_x$B$_2$C samples exhibit an enormous out-of-plane anisotropy. This unambiguously demonstrates that this anisotropy is due neither to the magnetic impurities (as those crystals have a density of impurities much smaller than the YNi$_2$B$_2$C crystal), nor to nonlocality (as it does not dissapear with increasing $x$). We also rule out the precence of surface barriers for ${\bf H}\bot c$ by performing minor hysterisis loops.

\section{EXPERIMENTAL DETAILS}

We report magnetization measurements, performed with a Quantum Design SQUID magnetometer, on YNi$_2$B$_2$C (Y-0), and Lu(Ni$_{1-x}$Co$_x$)$_2$B$_2$C with $x=0$ (Lu-0), $x=1.5 \%$ (Lu-1.5) and $x=3 \%$ (Lu-3) single crystals. The dimensions, $T_c$ and estimated $\ell$ for each sample are summarized in Table I. The Y-0 crystal is the same one that was previously investigated in Refs.\cite{song99a,basalplane,fourfold,physicaC,Fp}. The normal state magnetization follows a Curie law that indicates the presence of a very dilute distribution of localized magnetic moments. It corresponds to a rare-earth impurity content of $0.1$ at. $\%$ relative to Y, probably due to contaminants in the Y starting material\cite{basalplane}. The Lu(Ni$_{1-x}$Co$_x$)$_2$B$_2$C crystals, grown as described elsewhere\cite{cheon98}, show a much weaker Curie tail at low temperatures ($T < 100 K$) which might arise, for example, from a $0.001 \%$ magnetic impurities of Gd in the Lu site\cite{cheon98}. Isothermal magnetization loops in the superconducting mixed state were measured, and the critical current density $J_c$ was then calculated using the Bean's critical state model\cite{evidence,anomalous}.

\section{RESULTS AND DISCUSIONS}

\subsection{Reorientation phase transition}

\begin{figure}[htb]
\centering
\includegraphics[angle=0,width=90mm]{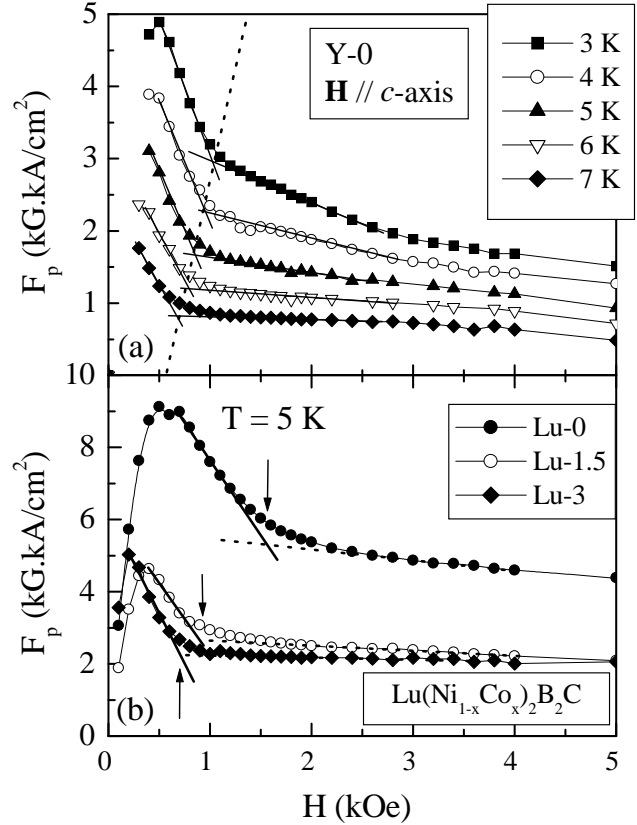}
\caption[]{{\small Field dependence of the pinning force density for $\bf H \parallel c$-axis for (a) Y-0 sample at several temperatures and (b) Lu-0, Lu-1.5 and Lu-3 samples at $T=5K$.}}
\label{FpvsH}
\end{figure}

Figure \ref{FpvsH}(a) shows the pinning force density $F_p=\left| \bf {J_c \times B} \right|$ for the \mbox{Y-0} sample as a function of the applied field for $\bf H \parallel c$-axis at several $T$. We observe that at low fields $F_p(H)$ decreases strongly with increasing $H$, but above a field $H^* \sim 1.2 kOe$ the field dependence becomes much weaker. In a recent work\cite{Fp} we showed that this "kink" in $F_p(H)$ at $H^*$ is a signature of the reorientation phase transition in the FLL. We based this claim in several facts. First, the position of the kink for $\bf H \parallel c$ coincides with the value of $H_1$. Second, $H^*$ is rather insensitive to the field orientation, in agreement with the behavior of $H_1$ observed in SANS experiments\cite{mckpaul98}. Finally, $H_1/H_{c1}$ is an increasing function of $T$, as predicted by Knigavko et al.\cite{knigavko}.

\begin{table*}[ht]
\begin{center}
\caption[]{volume, thickness, superconducting critical temperature and electronic mean free path for the investigated samples.}
\begin{tabular}{crrrc}                                    
\hline
$sample$ & \multicolumn{1}{c}{$V(mm^3)$} & \multicolumn{1}{c}{$t(mm)$} & \multicolumn{1}{c}{$T_c(K)$} & \multicolumn{1}{c}{$l(\AA)$} \\ 
\hline
$Y-0$ & $2.8$ & $0.5$ & $15.1$ & $300$ \\
$Lu-0$ & $2.5$ & $0.3$ & $15.7$ & $270$ \\
$Lu-1.5$ & $1.2$ & $0.4$ & $14.9$ & $100$ \\
$Lu-3$ & $0.44$ & $0.2$ & $14.1$ & $70$ \\
\hline
\end{tabular}
\end{center}
\end{table*}

We now expand the previous study and analyze the behavior of the field $H^*$ when the nonlocality is reduced. In the Kogan-Gurevich description\cite{kogan96,kogan97a} the strength of the nonlocal perturbations is parametrized by a new characteristic distance, the nonlocality radius $\rho(T,\ell)$: the weaker the nonlocality effects, the smaller $\rho(T,\ell)$. This means that $\rho(T,\ell)$ decreases with increasing $T$ or decreasing $\ell$. Thus, if this model is correct, an increase in $T$ or a decrease in $\ell$ should produce qualitatively similar effects on the nonlocality-induced transitions $H_1$ and $H_2$.

A typical way to reduce $\ell$ is to introduce impurities. It has been shown that the increase of $x$ in Lu(Ni$_{1-x}$Co$_x$)$_2$B$_2$C crystals decreases $\ell$ without significantly increasing $J_c$\cite{gammel99a}. Recent results confirmed that the analogy between $T$ and $\ell$ is indeed valid in this material in the case of $H_2$. First, Gammel et al.\cite{gammel99a} showed that $H_2$ increases as $\ell$ is reduced by increasing $x$. Later on, Eskildsen et al.\cite{eskildsen01} showed that $H_2(T)$ also rises as $T$ increases. 

\begin{figure}[htb]
\centering
\includegraphics[angle=0,width=90mm]{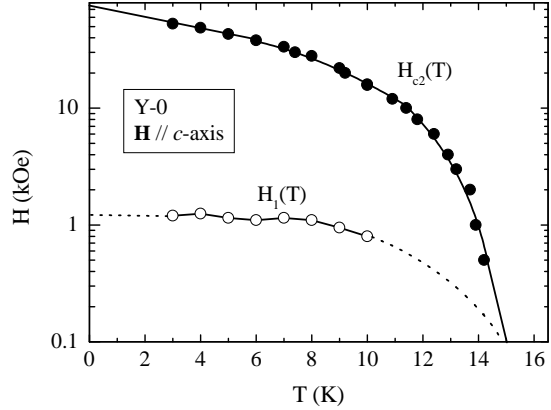}
\caption[]{{\small Temperature dependence of the upper critical field and the structural reorientation transition field for the Y-0 sample when ${\bf H}\parallel c$.}} 
\label{phasediagram}
\end{figure}

If $H^*$ is indeed a signature of $H_1$, we should also find a correlation between its $T$ and $\ell$ dependencies. In Figure \ref{phasediagram} we plot $H^*(T)$, together with $H_{c2}(T)$. We observe that $H^*$ is almost constant at low $T$ and, unlike the rhombic to square second order transition $H_2(T)$, it slightly decreases with $T$ at higher temperatures. 

To determine the $\ell$ dependence of $H^*$ we performed measurements on the Lu-0, Lu-1.5 and Lu-3 samples at $T = 5 K$ for $\bf H \parallel c$ (estimates of $\ell$ are given in Table I). The results are shown in Figure \ref{FpvsH}(b), where we plotted $F_p(H)$. We observe that the kink in $F_p$ (indicated by the arrows) is still visible in the field range where the transition $H_1$ should appear, and that it shifts to lower fields with increasing $x$, in agreement with the $T$ dependence. In other words, the analogy between $x$ and $T$ is satisfied for $H^*$, thus confirming that the kink arises from nonlocality. 

Another interesting fact is that the Lu-0 sample has a larger $F_p$ than the Y-0 sample at the same $T$, even though it has a lower density of magnetic impurities. This indicates that the magnetic moments in the Y-0 are not the relevant pins for the flux lines when ${\bf H}\parallel c$.

\subsection{Out-of-plane anisotropy}

\begin{figure}[htb]
\centering
\includegraphics[angle=0,width=120mm]{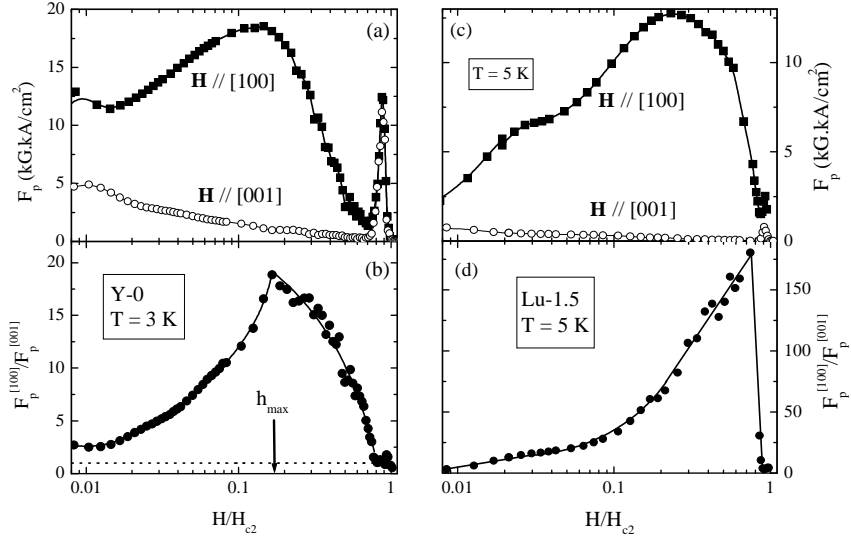}
\caption[]{{\small (a) Pinning force density for the samples Y-0 at $T = 3 K$ and (c) Lu-1.5 at $T = 5 K$ for ${\bf H}\parallel c$ and ${\bf H}\bot c$. (b) and (d) show the out of plane anisotropy $\Gamma$ corresponding to (a) and (c) respectively.}} 
\label{outofplane}
\end{figure}

We now turn to the pinning properties for ${\bf H} \bot c$. Figure \ref{outofplane}(a) shows $F_p$ at $T = 3K$ for the Y-0 sample for ${\bf H}\parallel [100]$, as a function of the reduced field $h~=~H/H_{c2}$. The ${\bf H} \parallel [001]$ data, already shown in fig. \ref{FpvsH}(a), is included for comparison. It is evident that the behavior for ${\bf H}\parallel [100]$ is very different from that observed for ${\bf H} \parallel [001]$. First we note that a broad maximum develops at intermediate fields $h_{max} \sim 0.15$. Second, $F_p$ is much larger than for ${\bf H} \parallel c$ over most of the field range. The maximum of this out-of-plane anisotropy, $\Gamma = F_p[100]/F_p[001]$, is $\sim 20$ as shown in Figure \ref{outofplane}(b). At higher $T$ the overall behavior and the $\Gamma$ values are similar. Finally, for ${\bf H} \parallel [110]$ (not shown), we found that $F_p$ is slightly but systematically smaller ($\sim 12 \%$) than for ${\bf H} \parallel [100]$.

Hereafter we will focus on the origin of the difference in $F_p$ between the $c$-axis and the $ab$-plane. As we pointed out in ref.\cite{Fp}, the large out-of-plane $F_p$ anisotropy sharply contrasts with the very small ($< 10 \%$) mass anisotropy\cite{basalplane,fourfold,kogan99a}. Therefore, explanations based on the anisotropic scaling frequently used in high $T_c$ superconductors\cite{blatter92} are ruled out.

As we mentioned above, the Y-0 crystal contains a small amount of magnetic impurities\cite{basalplane}, which align preferentially along the ab-plane and thus are a potential source of the $F_p$ anisotropy. However, we had previously argued\cite{Fp} that this was unlikely. Indeed, since the alignment of these localized moments increases with $H$, pinning should become more directional as the field increases and thus $\Gamma$ should increase monotonically with $H$. This is not in agreement with the data in Fig.\ref{outofplane}(b), where $\Gamma$ first grows with $h$, maximizes at $h \sim h_{max}$ and decreases again. In particular, $F_p$ is almost isotropic at the peak effect (see dotted line which corresponds to $\Gamma=1$). Thus, although we were not able to totally rule out magnetic pinning on crystal Y-0, the data suggested that this was not the case.

Conclusive evidence that the localized magnetic moments are not responsible for the large $\Gamma$ comes from the persistence of this effect in the samples Lu-0, Lu-1.5 and Lu-3, which have a magnetic impurity content two order of magnitude lower than the Y-0 crystal. In Figure~\ref{outofplane}(c) we show $F_p(h)$ for the Lu-1.5 for ${\bf H} \parallel c$ and ${\bf H} \bot c$, at $T = 5 K$, and in Figure.~\ref{outofplane}(d) we plotted the corresponding $\Gamma(h)$. We observe that the out-of-plane anisotropy is even larger than in the Y-0 sample. In contrast to the behavior observed in the Y-0, $\Gamma$ in the Lu-1.5 increases monotonically with $h$ up to near the peak effect region, where it suddenly drops approaching to the isotropic limit. Measurements on the Lu-3 crystal at several $T$ show a similar behavior and exhibit a $\Gamma$ larger than in the Y-0.

We had also previously argued\cite{Fp} that $\Gamma$ in the Y-0 sample seems too large to be ascribed to nonlocality, which should appear as a perturbatively small effect. This conclusion is unambiguously confirmed by the presence of even larger $\Gamma$ values in the doped samples Lu-1.5 and Lu-3, where nonlocal effects are strongly suppressed.

\begin{figure}[htb]
\centering
\includegraphics[angle=0,width=90mm]{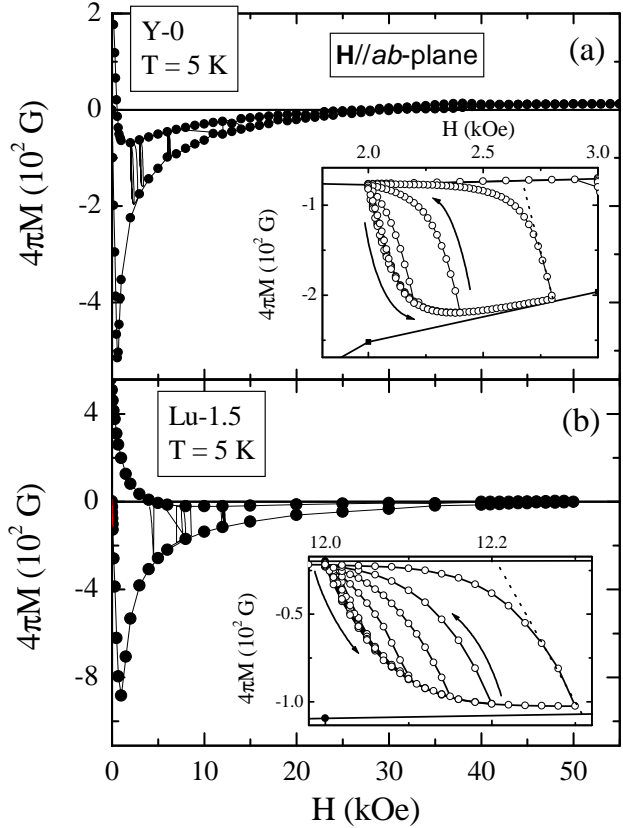}
\caption[]{{\small main panel: hysterisis loop for ${\bf H}\bot c$ at $T = 5 K$ for (a) Y-0 sample and (b) Lu-1.5 sample. The insets show a blow up of the minor hysterisis loops corresponding to the full loop showed in the main panel. The dotted lines show the behavior expected for surface barriers.}} 
\label{minorloops}
\end{figure}

The existence of significant surface barriers for ${\bf H} \bot c$, have also been ruled out by performing minor hysteresis loops with ${\bf H} \parallel ab$ at several $T$ and $H$. Examples for the Y-0 and Lu-1.5 samples are shown in Figure~\ref{minorloops}. If hysteresis were due to surface barriers no flux changes would occur in the bulk while $H$ is changing from one branch of the main loop to the other one, hence the data of the minor loop connecting the lower and upper branches would be Meissner-like straight lines\cite{ullmaier75}. In contrast, in the case of bulk pinning, the lines connecting both branches are curved (parabolic in the simplest Bean model for an infinite slab) just as we observe in the insets of Figure \ref{minorloops}(a) and (b). 

Moreover, the $F_p$ calculated from these minor loops assuming only bulk pinning are in good agreement with those obtained from the main loops (Fig. \ref{outofplane}). Indeed, according to the isotropic Bean model the critical state profile can be completely inverted if we change the external field in $\Delta H= 2 (4\pi \Delta M)/g$, where $g = 1-t/3L$, and $t$ and $L$ are the shortest and largest sample dimensions perpendicular to the field direction, respectively. From the inset of figure~\ref{minorloops}(a) we find that the width of the hysterisis loop is $4\pi \Delta M \approx 150 G$. For this sample $g \sim 0.9$, thus it is expected that the connecting line between the two branches of the loop, reaches the upper branch, when the field is decreased about $\Delta H \approx 330 G$. Inspection of figure~\ref{minorloops}(a) shows that $\Delta H \approx 500 G$, in reasonably good agreement given the simplicity of the model. The same analysis on the data of the inset of figure~\ref{minorloops}(b), leds to a similar conclusion.

\section{CONCLUSIONS}
Non-magnetic borocarbides provide a very attractive ground to study the limit of very low pinning forces. The high quality of the crystals have the double benefit of producing large  mean free paths, resulting in measurable non-local effects, and a very dilute distribution of vortex pinning centers. The combination of both factors allows the observation of the influence of nonlocality on vortex pinning. This is a rather unexplored field, and further studies, both experimental and theoretical, are needed. On the other hand, the origin of the large out-of-plane anisotropy in $F_p$ is unclear and deserves further investigation. We have conclusively ruled out explanations based on the mass anisotropy, pinning by magnetic impurities, non-local effects and surface barriers. A simple explanation for it could be the presence of some still unidentified anisotropic pinning centers, such as planar defects.

\section{ACKNOWLEDGMENTS}

Work partially supported by ANPCyT, Argentina, PICT 97 No.01120, and CONICET PIP 4207. A.V.S. would like to thank to CONICET for financial support. Research sponsored by the U.S. Department of Energy under contract DE-AC05-00OR22725 with the Oak Ridge National Laboratory, managed by UT-Battelle, LLC.


\begin{thebibliography}{99}

\bibitem{mazumdar}
C.~Mazumdar {\it et al.}, {\em Solid State Commun.} {\bf 87}, 413 (1993).

\bibitem{nagarajan}
R.~Nagarajan, C.~Mazumdar, Z. Hossain, S. K. Dhar, K. V. Gopalakrishnan,
L. C. Gupta, C. Godart, B. D. Padalia, and R. Vijayaraghavan, {\em Phys. 
Rev. Lett.} {\bf 72}, 274 (1994).

\bibitem{cava94a}
R.J.~Cava, H. Takagi, H. W. Zandbergen, J. J. Krajewski, W. F. Peck, Jr., 
T. Siegrist, B. Batlogg, R. B. van Dover, R. J. Felder, K. Mizuhashi, J. O. 
Lee, H. Eisaki, and S. Uchida, {\em Nature (London)} {\bf 367}, 252 (1994).

\bibitem{cava94b}
R.J.~Cava, H. Takagi, B. Batlogg, H. W. Zandbergen, J. J. Krajewski, 
W. F. Peck, Jr., R. B. van Dover, R. J. Felder, T. Siegrist, K. Mizuhashi, 
J. O. Lee, H. Eisaki, S. A. Carter, and S. Uchida, {\em Nature (London)} 
{\bf 367}, 146 (1994).

\bibitem{yaron96}
U. Yaron, P.L. Gammel, A.P. Ramirez, D.A. Huse, D.J.
Bishop, A.I. Goldman, C. Stassis, P.C. Canfield, K. Mortensen, and M.R.
Eskildsen, {\em Nature} {\bf 382}, 236 (1996).

\bibitem{Eisaki94}
H. Eisaki, H. Takagi, R. J. Cava, B. Batlogg, J. J.
Krajewski, W. F. Peck, Jr., K. Mizuhashi, J. O. Lee, and S. Uchida, {\em
Phys. Rev. B} {\bf 50}, 647 (1994).

\bibitem{Eskildsen98}
M. R. Eskildsen; K. Harada; P. L. Gammel; A. B.
Abrahamsen; N. H. Andersen; G. Ernst; A. P. Ramirez; D. J. Bishop, K.
Mortensen, D. G. Naugle, K. D. D. Rathnayaka, and P. C. Canfield, {\em Nature
} {\bf 393}, 242 (1998).

\bibitem{Gammel99}
P. L. Gammel, B. P. Barber, A. P. Ramirez, C. M. Varma,
D. J. Bishop, P. C. Canfield V. G. Kogan, M. R. Eskildsen, N. H. Andersen,
K. Mortensen, and K. Harada, {\em Phys. Rev. Lett.} {\bf 82}, 1756 (1999).

\bibitem{Norgaard00}
K. N\o rgaard, M. R. Eskildsen, and N. H. Andersen,
J. Jensen, P. Hedeg\aa rd, and S. N. Klausen, and P. C. Canfield, {\em Phys.
Rev. Lett.} {\bf 84}, 4982 (2000).

\bibitem{carter}
S.A.~Carter {\it et al.}, {\em Phys. Rev. B} {\bf 50}, 4216 (1994).

\bibitem{hong}
N.M.~Hong {\it et al.}, {\em Physica C} {\bf 226}, 85 (1994).

\bibitem{nohara}
M.~Nohara {\it et al.}, {\em J. Phys. Soc. Jpn.} {\bf 66}, 1888 (1997).

\bibitem{shulga}
S.V.~Shulga, S. L. Drechsler, G. Fuchs, K. H. Mueller,
K. Winzer, M. Heinecke, and K. Krug, {\em Phys. Rev. Lett.} {\bf 80}, 1730 (1998).

\bibitem{ovchinnikov00}
Y.N. Ovchinnikov and V.Z. Kresin, {\em Eur. Phys. J. B} {\bf 14}, 203-209 (2000).

\bibitem{metlushko97a}
V. Metlushko, U. Welp, A. E. Koshelev, I. Aranson, G. W.
Crabtree, and P. C. Canfield, {\em Phys. Rev. Lett.} {\bf 79}, 1738 (1997).

\bibitem{yethiraj97a}
M.~Yethiraj, D.McK.~Paul, C.V.~Tomy, and E.M.~Forgan, {\em Phys. Rev. Lett.} 
{\bf 78}, 4849 (1997).

\bibitem{mckpaul98}
D.McK.~Paul, C.V.~Tomy, C. M. Aegerter, R. Cubitt, S. H. Lloyd, E.M.~Forgan,
S. L. Lee, and M.~Yethiraj, 
{\em Phys. Rev. Lett.} {\bf 80}, 1517 (1998).

\bibitem{wang-maki}
G.~Wang and K.~Maki, {\em Phys. Rev. B} {\bf 58}, 6493 (1998).

\bibitem{kogan96}
V.G. Kogan, A. Gurevich, J.H. Cho, D.C. Johnston, Ming
Xu, J.R. Thompson and A. Martynovich, {\em Phys. Rev. B} {\bf 54}, 12386 (1996).

\bibitem{dewilde97}
Y.~De Wilde, M. Iavarone, U. Welp, V. Metlushko, A. E. Koshelev, I. Aranson, G. W.
Crabtree, and P. C. Canfield, {\em Phys. Rev. Lett.} {\bf 78}, 4273 (1997).

\bibitem{park-huse}
K.~Park and D.A.~Huse, {\em Phys. Rev. B} {\bf 58}, 9427 (1998).

\bibitem{kogan88a}  V.G. Kogan, M.M. Fang, and S. Mitra, {\em Phys. Rev. B} 
{\bf 38}, 11958 (1988).

\bibitem{song99a}  K.J. Song, J.R. Thompson, M.Yethiraj, D.K. Christen,
C.V. Tomy and D. McK. Paul, {\em Phys. Rev. B} {\bf 59}, R6620 (1999).

\bibitem{basalplane}
J. R. Thompson, A.V. Silhanek, L. Civale, K.J. Song, D. McK Paul and C.V. Tomy. {\em Phys. Rev. B} {\bf 64}, 24510 (2001).

\bibitem{kogan99a}  
V.G. Kogan, S.L. Bud\'{}ko, P.C. Canfield and P.
Miranovic, {\em Phys. Rev. B} {\bf 60}, R12577 (1999).

\bibitem{fourfold}  L. Civale, A.V. Silhanek, J.R. Thompson, K.J. Song,
C.V. Tomy and D.McK. Paul, {\em Phys. Rev. Lett.} {\bf 83}, 3920 (1999).

\bibitem{physicaC}
L. Civale, A.V. Silhanek, J.R. Thompson, K.J. Song,
C.V. Tomy and D.McK. Paul, {\em Physica C} {\bf 341-348}, 1299 (2000).

\bibitem{kogan97a} 
V.G. Kogan, M. Bullock, B. Harmon, P. Miranovi\'{c}, Lj. Dobrosavljevi\'{c}-Gruji\'{c}, P.L. Gammel, and D.J. Bishop, {\em Phys. Rev. B} {\bf 55}, R8693 (1997).

\bibitem{knigavko}
A. Knigavko, V.G. Kogan, B. Rosenstein and T.J. Yang, {\em Phys. Rev. B} {\bf 62}, 111 (2000)

\bibitem{eskildsen97b}
M.R.~Eskildsen, P.L.~Gammel, B.P.~Barber, U.~Yaron, A.P.~Ramirez, D.A.~Huse, D.J.~Bishop, C.~Bolle, C.M.~Lieber, S.~Oxx, S.~Sridhar, N.H.~Andersen, K.~Mortensen, and P. C. Canfield, {\em Phys. Rev. Lett.} {\bf 78}, 1968 (1997).

\bibitem{rosenstein99}
B. Rosenstein and A. Knigavko, {\it Phys. Rev. Lett.} {\bf 83}, 844 (1999).

\bibitem{Fp}
A.V. Silhanek, J. R. Thompson, L. Civale, D. McK Paul and C.V. Tomy. {\em Phys. Rev. B} {\bf 64}, 12512 (2001).

\bibitem{cheon98}  
K.O.~Cheon, I.R.~Fisher, V.G. Kogan, P.C. Canfield, P. Miranovic, and P.L. Gammel  {\em Phys. Rev. B} {\bf 58}, 6463 (1998).

\bibitem{evidence}
A.V. Silhanek, L. Civale, S. Candia, G. Nieva, G. Pasquini, H. Lanza, {\em Phys. Rev. B} {\bf 59}, 13620 (1999).

\bibitem{anomalous}
A.V. Silhanek, D. Niebieskikwiat, L. Civale, M. A. Avila, O. Billoni and D. Casa, {\em Phys. Rev. B} {\bf 60}, 13189 (1999).

\bibitem{gammel99a}
P. L. Gammel, D. J. Bishop, M. R. Eskildsen, K. Mortensen, N. H. Andersen, I.R.~Fisher, 
K.O.~Cheon, P. C. Canfield, and V. G. Kogan, {\it Phys. Rev. Lett.} {\bf 82}, 4082 (1999).

\bibitem{ullmaier75}
H. Ullmaier, {\it Irreversible Properties of Type II Superconductors} (Springer-Verlag, Berlin-Heidelberg-New York, 1975), p. 124.

\bibitem{eskildsen01}
M.R.~Eskildsen, A.B.~Abrahamsen, V.G.~Kogan, P.L.~Gammel, K.~Mortensen, N.H.~Andersen, and P. C. Canfield, {\em Phys. Rev. Lett.} {\bf 86}, 5148 (2001).

\bibitem{blatter92}
G. Blatter, V.B. Geshkenbein and A.I. Larkin. Phys. Rev. Lett. {\bf 68}, 875 (1992).

\end{thebibliography}
\end{document}